\documentclass[aps,pre,twocolumn,superscriptaddress,floatfix]{revtex4-1}
\usepackage{graphicx,amsmath,subfigure,epsfig,psfrag,verbatim,color}
\usepackage{epstopdf}
\begin{document}
\title{Distinguishing XY from Ising Electron Nematics}

\author{S. Basak}
\affiliation{Department of Physics, Purdue University, West Lafayette, IN 47907, USA}
\author{E.~W. Carlson}
\email{ewcarlson@purdue.edu}
\affiliation{Department of Physics, Purdue University, West Lafayette, IN 47907, USA}

\date{\today}

\begin{abstract}
At low temperatures in ultraclean 
GaAs-AlGaAs 
heterojunctions, high fractional Landau levels break rotational symmetry,
leading to increasingly anisotropic transport properties as temperature is lowered below 
$\sim$150mK. 
While the onset of transport anisotropy is well described by an XY model of an electron nematic 
in the presence of a weak uniform symmetry-breaking term, the low temperature behavior deviates significantly from this model.  
We find that inclusion of interactions between the electron nematic and the underlying crystalline lattice 
in the form 
of a 4-fold symmetry breaking term is sufficient to describe the entire temperature
dependence of the transport anisotropy. This implies that the quantum Hall electron nematic is in the
Ising universality class.  We propose new experimental tests 
that can distinguish whether any two-dimensional electron nematic is in the XY or Ising universality class. 
\end{abstract}

\maketitle

\label{sxn:introduction}

Strong electron correlations can drive systems into a variety of novel electronic phases of matter.
Electronic liquid crystals\cite{Kivelson1998,Fradkin-Kivelson-1999-PhysRevB.59.8065,RevModPhys.75.1201} form when electronic degrees of freedom partially break the symmetries of the host crystal.
Like their molecular counterparts,  electron nematic phases break rotational symmetry,
while retaining liquidity.  Such oriented electronic liquids have been observed in a variety of 
systems, including strontium ruthenates\cite{Borzi214}, iron superconductors\cite{PhysRevB.77.224509,PhysRevB.78.020501,Mazin2009},
cuprate superconductors\cite{PhysRevLett.88.137005,Hinkov597}, and high fractional Landau levels.   
The key signature in the quantum Hall regime is a pronounced 
transport anisotropy that develops at low temperature.\cite{annurev-conmatphys-070909-103925,Lilly-1999-PhysRevLett.82.394,Du-1999,PhysRevX.4.041050, PhysRevB.93.205401}

At high fractional Landau levels,
uniform quantum Hall phases are unstable to the formation
of stripe and bubble phases, with the stripe phases being
preferred near high half-filling.\cite{Fogler:1996kp}
Several stripe phases are possible, 
including (insulating) stripe crystals, 
as well as (compressible) electronic liquid crystal phases like
nematic or smectic.\cite{Fradkin-Kivelson-1999-PhysRevB.59.8065}
The quantum Hall state in GaAs-AlGaAs  heterojunctions
at filling $\nu = 9/2$\cite{Lilly-1999-PhysRevLett.82.394,Du-1999} 
has been identified as a nematic.\cite{Nho-2000-PhysRevLett.84.1982}
Fradkin {\em et al.}\cite{Nho-2000-PhysRevLett.84.1982} 
developed an order parameter theory of the nematic to describe 
the temperature evolution of the resistivity anisotropy as it develops.
Using symmetry to map the resistivity anisotropy to 
the nematic order parameter, they showed that the temperature
evolution of the resistivity anisotropy in the $\nu=9/2$ state
is well described by a classical 2D XY model, with a weak uniform symmetry-breaking
term, through the onset of the resistivity anisotropy as temperature is lowered
below $\sim$150mK, with deviations from the theory beginning below $\sim$55mK. 
This model places the transition in the BKT universality class\cite{Berezinskii:1971,Kosterlitz:2001}.

One difficulty with this identification is that a true BKT transition does not
break symmetry, and in fact in that model long-range order
of a nematic is forbidden at finite temperature.
However, as stressed in Ref.~\onlinecite{Nho-2000-PhysRevLett.84.1982},
the nematic susceptibility is sufficiently
strong in the BKT phase that 
net nematicity can develop anyway
in the presence of even a weak uniform orienting field.
Note that without the development of net nematicity,
the resistivity anisotropy would be zero.

\begin{figure}
\centering
\includegraphics[width=1.05\columnwidth]{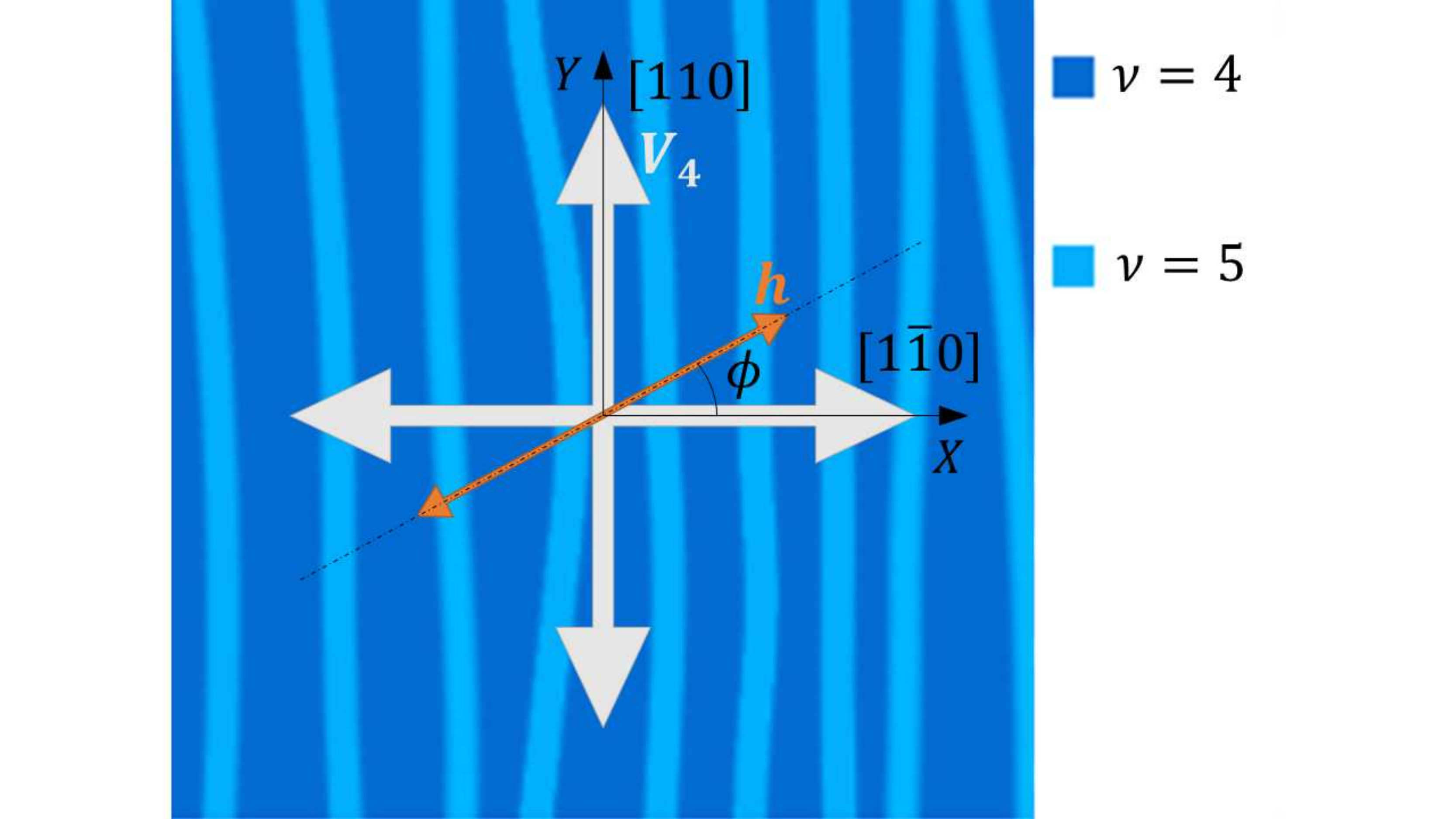}
\caption{The 4 gray arrows  represent
the directions in which the crystal field term ($V$) is maximum. The orange double headed arrow gives the orientation of the h-field with respect to the crystal field. 
The resistivities $\rho_{xx}$ and $\rho_{yy}$ are typically measured along the crystallographic directions
 [$ 1 \bar{1} 0 $] and [$ 1 1 0 $] , respectively.} \label{fig:phi_angle}
\end{figure}

Here, we propose a model of the quantum Hall nematic
which solves both the problem of the deviation of the low
temperature resistivity anisotropy data from the order parameter theory,
as well as the issue of long-range order.  
The nematic order parameter is a headless vector, which depends on overall orientation but not the direction. 
(That is, it is symmetric with respect to a 180$^o$ rotation.) 
If the interaction between the electron nematic and the host crystal
is sufficiently weak, the nematic is free to form in any direction,
and an XY model of the development of nematicity is appropriate.\cite{Nho-2000-PhysRevLett.84.1982}
We first consider this case, but in the presence of 
lattice effects which ultimately at low temperature lock
the nematic to a crystalline axis:  
\begin{eqnarray}
H=&-&J\sum_{\langle i,j \rangle}\cos{(2(\theta_{i}-\theta_{j}))}-h\sum_{i}\cos{(2(\theta_{i}-\phi))} \nonumber \\
&-& V\sum_{i}\cos{(4\theta_{i})},
\label{eqn:model_V4}
\end{eqnarray}
where J gives the interaction between neighboring regions, h is an orienting field, 
and the $V$ term captures the four-fold symmetry of the underlying lattice. The angle the orienting field $h$ makes with the crystal field is $\phi$ (Fig. \ref{fig:phi_angle}). 
The net orienting field $h$ can include intrinsic orienting effects, or be tuned via, {\em e.g.}, an applied orienting field
such as an in-plane magnetic field or strain, among other things.\cite{Carlson2011,KenBCooperThesis}
The ``nematicity'' (order parameter of the nematic) in this model is $\mathcal{N} = \left< e^{2 i \theta} \right>$.\cite{Nho-2000-PhysRevLett.84.1982}
Because the (normalized) macroscopic transport anisotropy $\rho_a$ transforms under rotations
in the same way as the nematicity, the two are related as
$\rho_a \equiv \big[(r+1)/(r-1) \big] \big( \rho_{xx} - \rho_{yy} \big)  / \big( \rho_{xx} + \rho_{yy} \big) = f(\mathcal{N})$
where $f(\mathcal{N})$ is an odd function of $\mathcal{N}$,
and $r \equiv \rho_{xx}(\mathcal{N} \rightarrow 1)/\rho_{yy}(\mathcal{N} \rightarrow 1)$
is what the ratio of macroscopic resistivities would be in a fully oriented state.  
For small $\mathcal{N}$, $f(\mathcal{N}) =  \mathcal{N}$.\cite{Nho-2000-PhysRevLett.84.1982,Carlson:2006}

\begin{figure}
\centering
\subfigure[\ XY model with moderate 4-fold symmetry breaking term V, Eqn.~\ref{eqn:model_V4}]{\label{fig:XY-nematic}\includegraphics[width=0.45\textwidth]{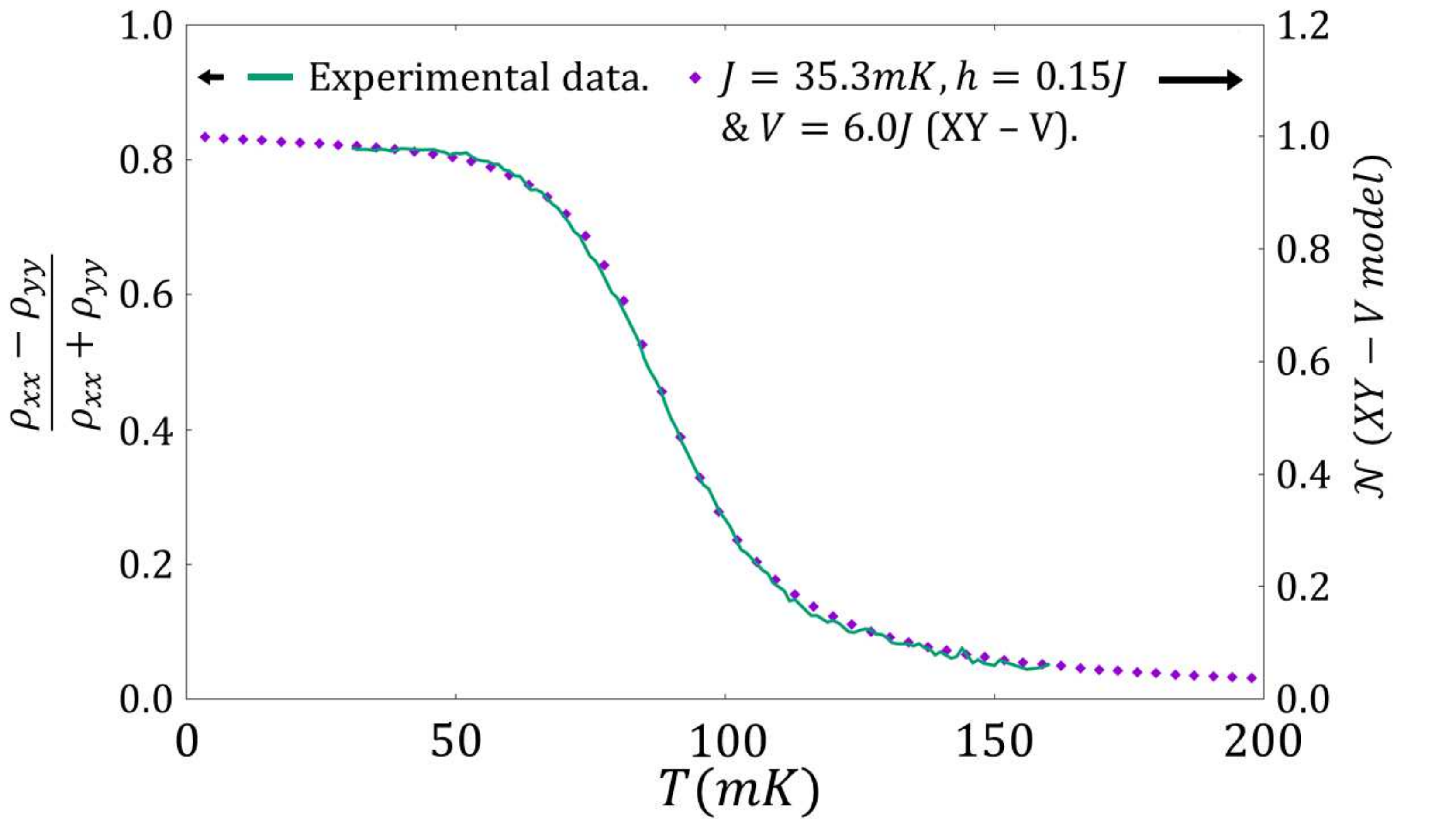}}
\subfigure[\ Ising Model, Eqn.~\ref{eqn:model_ising}]{\label{fig:ising-nematic}\includegraphics[width=0.45\textwidth]{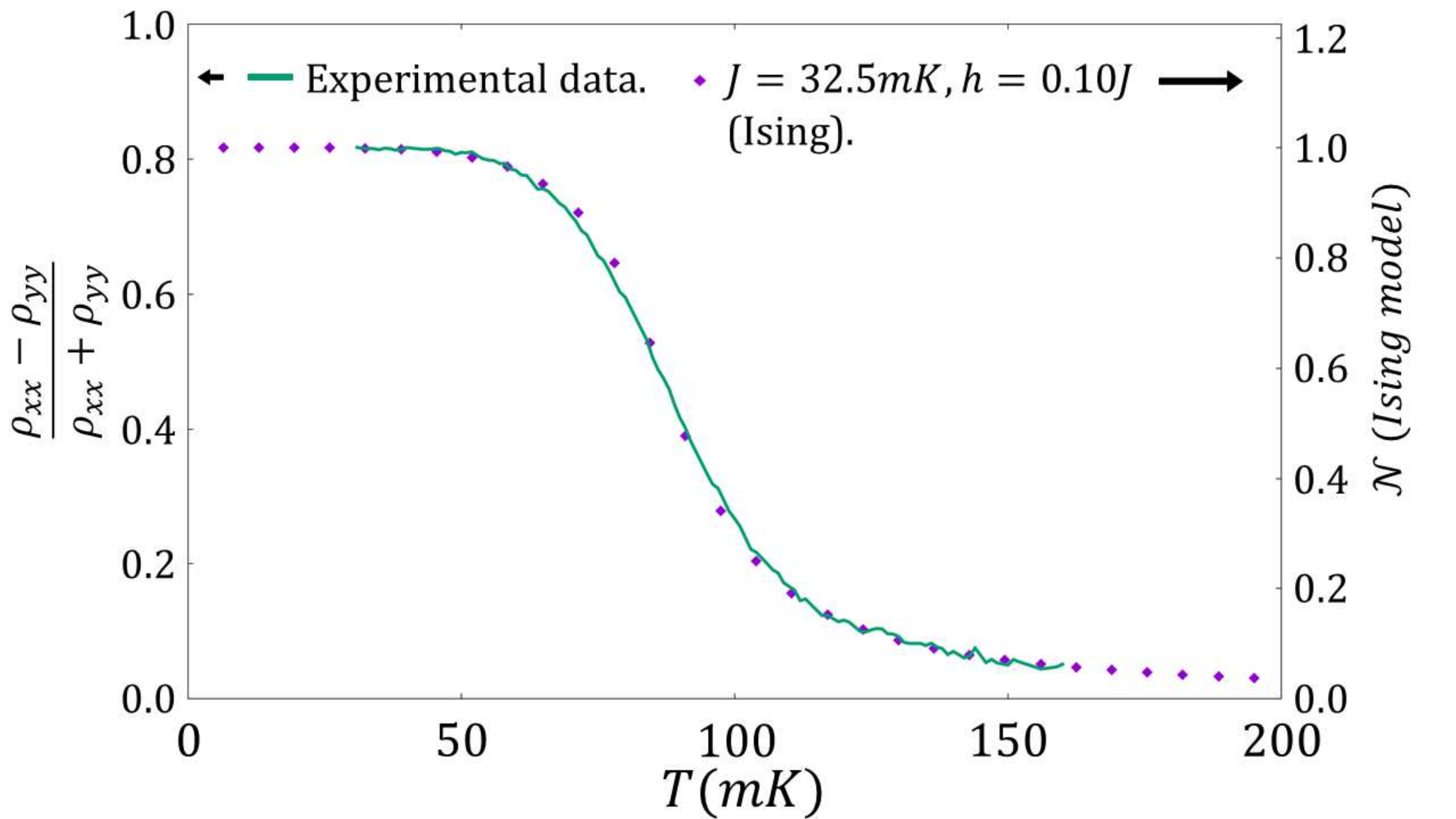}}
\caption{Monte Carlo simulations (purple dot) on a lattice of 100x100 sites,
compared to experimental data (green line) of resistivity anisotropy 
$\frac{\rho_{xx}-\rho_{yy}}{\rho_{xx}+\rho_{yy}}$ from Lilly {\em et al.}\cite{Lilly-1999-PhysRevLett.82.394}. 
The theoretical comparison is to: 
(a) an XY model with a moderate four-fold symmetry breaking field $V$ and uniform orienting field $h$, and 
(b) an Ising model with uniform orienting field $h$.  
Note that within an XY description, a moderate 4-fold symmetry breaking term $V \ne 0$ is required to 
capture the low temperature dependence of the resistivity anisotropy, 
which changes the universality class of the electron nematic from XY to Ising.    
The resistivity anisotropy  $\frac{\rho_{xx}-\rho_{yy}}{\rho_{xx}+\rho_{yy}}$ is from the experimental data of Lilly {\em et al.}\cite{Lilly-1999-PhysRevLett.82.394}. 
}
\label{fig:match_with_experiment}
\end{figure}

\begin{figure}
\centering
\includegraphics[width=1.0\columnwidth]{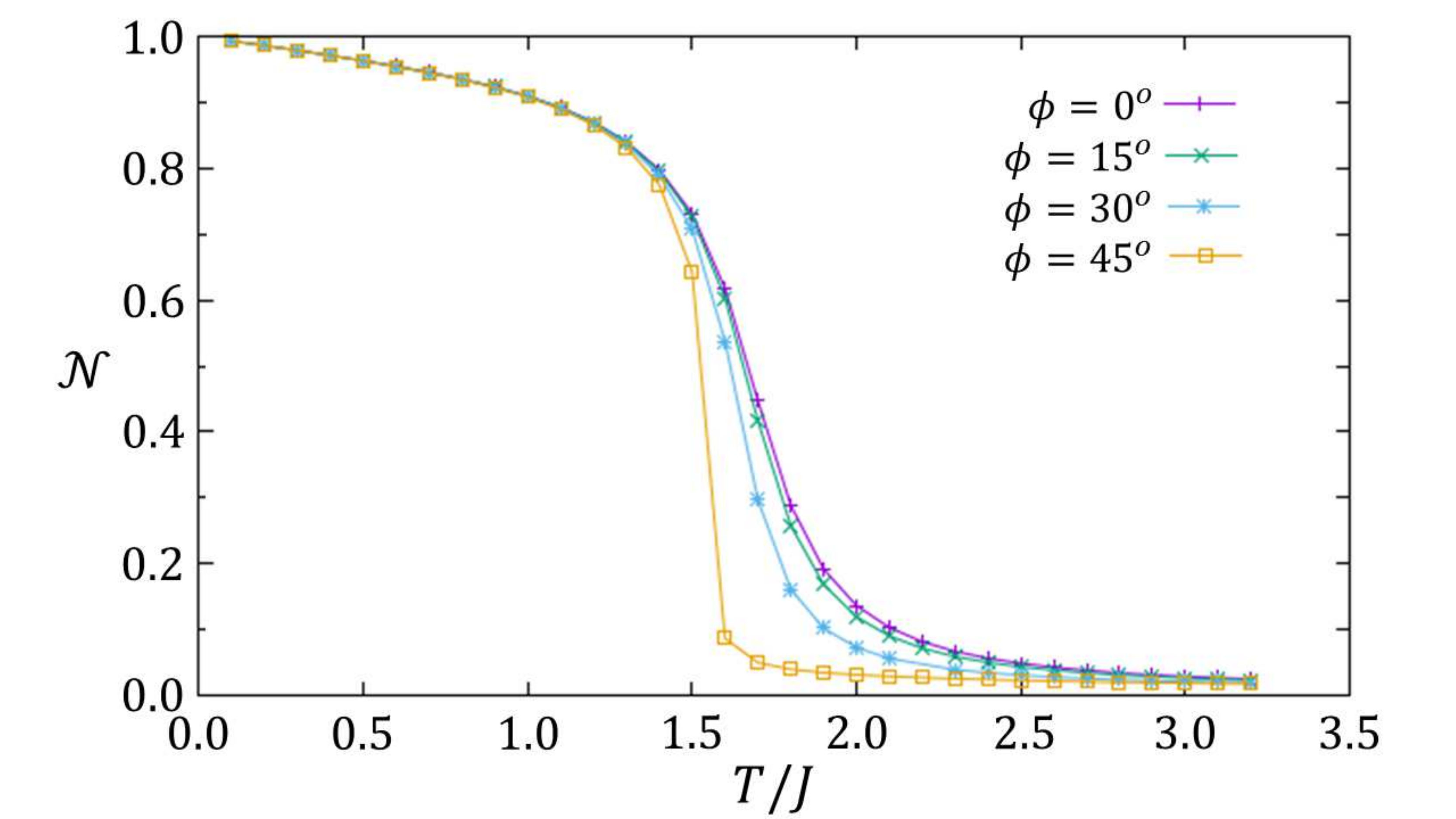}
\caption{Monte Carlo simulations on a lattice of 100x100 sites of an XY model with 
4-fold symmetry breaking term, for various angles $\phi$ of the orienting field $h$. 
This shows the increasing steepness of the nematic-to-isotropic transition when $\phi$ goes from $0^o\rightarrow 15^o\rightarrow 30^o\rightarrow 45^o$. All of these simulation are done with $h=0.05J$ and $V=1.0J$. } \label{fig:changing_phi}
\end{figure}

Results for the model of Eqn.~\ref{eqn:model_V4} are shown 
in Fig.~\ref{fig:XY-nematic}.   
As shown in Fig.~2 of  Ref.~\onlinecite{Nho-2000-PhysRevLett.84.1982}, the experimental data of Ref.~\onlinecite{Lilly-1999-PhysRevLett.82.394} can be 
matched reasonably well for $T \gtrsim 55$mK
in the presence of a weak uniform orienting field $h$ and $V=0$,
but with significant deviation below $55$mK. 
We find that the entire temperature evolution can be captured 
in the presence of both nonzero $h$ and nonzero $V$,
as shown in Fig.~\ref{fig:XY-nematic}.  
In the Figure, we use   uniform orientational field $h=.15J$ along 
with  four-fold symmetry breaking term $V=6J$, and $J=35.3$mK.  
Smaller values of $V$ have too steep of a  slope at low temperature.  
For larger values of $V$, the higher temperature behavior ($100 - 150$mK) 
can no longer be captured.  
For the parameters of Fig.~\ref{fig:XY-nematic}, the absolute strength of the interaction $J$
is about half that of Ref.~\onlinecite{Nho-2000-PhysRevLett.84.1982}.
Because the value of $V$ that we use is not small with respect to $J$, the universality class
of the transition is now Ising, not XY.  
For a pure XY model with $h=0$ and $V=0$, the transition temperature
is $T_{\rm KT} = .89J$.\cite{Kosterlitz:2001}, but 
in Fig.~\ref{fig:XY-nematic} the onset of nematicity is happening closer to the
(2D) Ising transition temperature of $T_c = 2.27J$,
consistent with this shift of universality class. 
(See Supplementary Information for results with other values of parameters.)

The effect of rotating the uniform orienting field $h$ away from a crystalline axis 
is explored in Fig.~\ref{fig:changing_phi},
where the angle $\phi$ between $h$ and the crystalline axes is varied.
Note that up until $\phi \approx 30^o$ the impact on the
temperature evolution is negligible.
However, at the high symmetry point $\phi = 45^o$,
there is a true symmetry breaking transition,
and the temperature onset is quite sudden.  

\label{sxn:Ising-model}
We have found that within an XY description (Eqn.~\ref{eqn:model_V4}), moderate values of $V/J$ are required to capture the
entire temperature dependence of the resistivity anisotropy in the quantum Hall nematic at $\nu = 9/2$. 
This naturally leads to the question of how well a simple Ising model can account for the data:
\begin{equation}
H=-J\sum_{\langle i,j \rangle}\sigma_{i} \sigma_{j}-h\sum_{j}  \sigma_{i}~.
\label{eqn:model_ising}
\end{equation}
Here, we make the assumption that the electron nematic, once it develops,
tends to lock to a favorable lattice direction.
In a crystal with 4-fold rotational symmetry, because the director of the
nematic is a headless vector, 
the order parameter of the nematic is explicitly in the Ising universality class
with the two possible orientations of the nematic being mapped to $\sigma = \pm 1$.  
A uniform orienting field (whether intrinsic or applied) is modeled by $h$.

The nematic order parameter in this case is $\mathcal{N} = (1/N)\sum_i \sigma_i$.
As in the case of an XY model of an electron nematic, 
the (normalized) macroscopic resistivity anisotropy $\rho_a$ maps to the 
macroscopic order parameter in the Ising description as 
$\rho_a = g(\mathcal{N})$ where $g(\mathcal{N})$ is an odd function of 
$\mathcal{N}$ and to first order in $\mathcal{N}$, $g = f$.

Note that for $h=0$, 
the magnetization for 2D-Ising model is 
\begin{equation}
M(T, h=0)=\bigg[1-sinh^{-4}\bigg(\frac{2J}{T}\bigg)\bigg]^{\frac{1}{8}}\ ,\ T < T_c ~.
\end{equation}
Unlike the XY model where the low temperature 
nematicity can only develop for nonzero $h$ and 
is linear in temperature $T$,
the Ising magnetization can develop even with $h=0$ and 
is flat at low temperatures.
Our comparison of the experimental resistivity anisotropy to an Ising model is 
shown in Fig.\ref{fig:ising-nematic}.
We find that the data can be well described throughout the entire temperature
range within a simple Ising model, 
with $ J = 32.5 mK $ and $ h = 0.1 J $. 

\label{sxn:Discussion}

Remarkably, we find that the entire temperature range of the resistivity anisotropy $\rho_a$ 
can be captured quite well within an Ising model in the presence of a weak
uniform orienting field.	
Within this context, the low temperature saturation of 
$\big( \rho_{xx} - \rho_{yy} \big)  / \big( \rho_{xx} + \rho_{yy} \big) $
to a value $ \approx .818 \ne 1$ 
could have several origins:\cite{Nho-2000-PhysRevLett.84.1982,Carlson:2006}
(i) Taken at face value, the saturation implies that the
bare ``nematogens'' represented by each Ising variable
have an intrinsic resistivity anisotropy
which persists down to the lowest temperatures,
$r = \rho_{xx}/\rho_{yy} \approx 10.$ 
This could be attributable to  quantum fluctuations
within a bare nematogen.  
(ii) Similar saturation effects could also arise from 
even a small amount of quenched disorder, since the
critical (random field type) disorder strength is zero
in a two-dimensional Ising model.
(iii)  Nonlinear terms in the function $g(\mathcal{N})$ 
can lead to $g \ne 1$ as $\mathcal{N} \rightarrow 1$ at low temperature.

\begin{figure}
\centering
\includegraphics[width=0.45\textwidth]{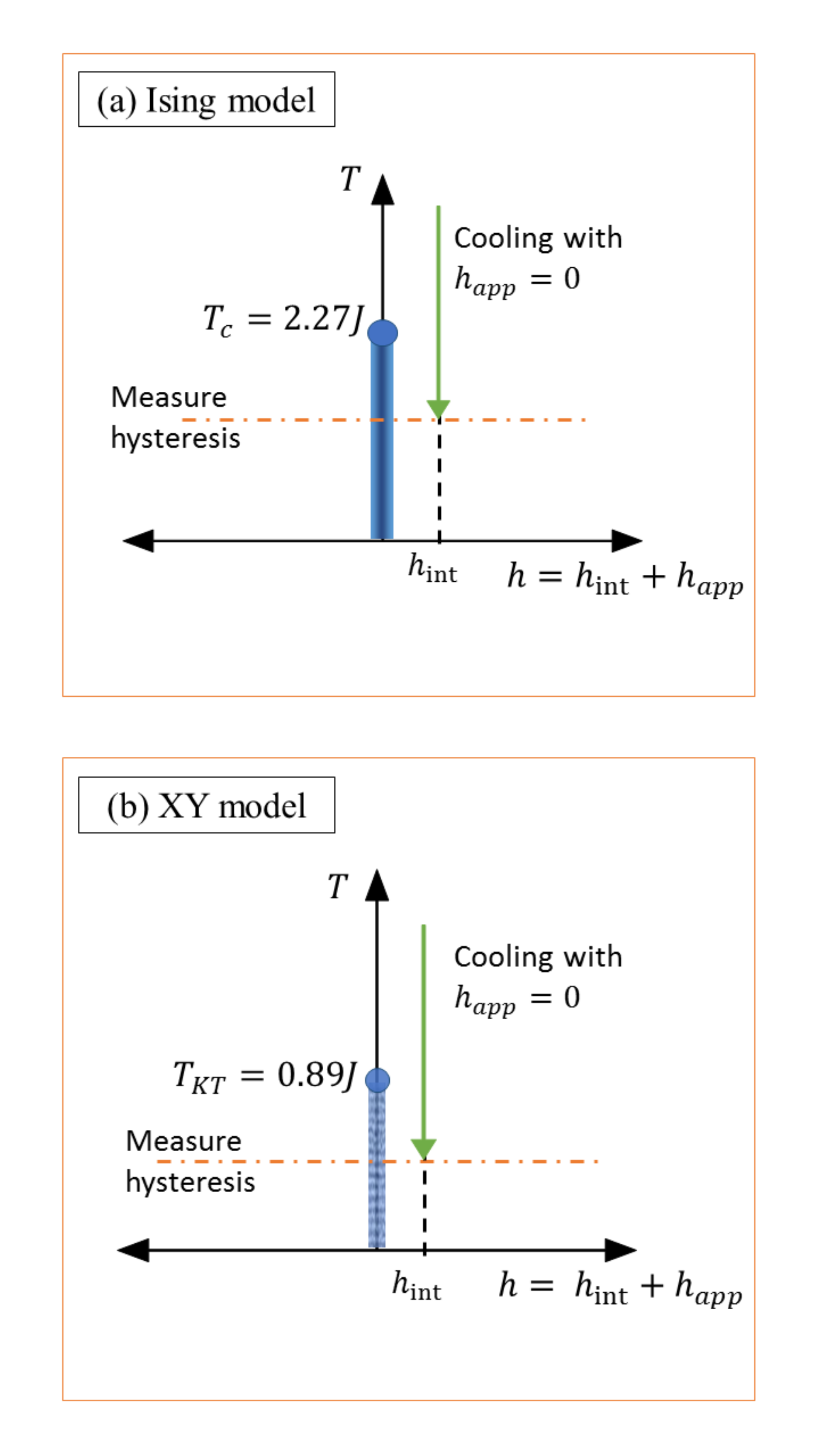}
\caption{Equilibrium phase diagram for (a) two-dimensional Ising model and (b) two-dimensional XY model.
In both cases, a low temperature phase transition occurs only for orienting field $h=0$.
In the Ising case, the low temperature phase has long-range nematic order, and in the 
XY case the low temperature phase only has topological order but no long-range nematic order. 
The experimental hysteresis test we propose begins by (i) cooling (green arrow)
with or without applied field $h_{app}$, followed by
(ii) sweeping the orienting field $h_{app}$ 
so as to move the system back and forth across the 
low temperature phase (orange dotted line).  
Refer to Fig.~\ref{fig:XY_vs_Ising-hysteresis}
for the experimental prediction of the response
of the nematicity $\mathcal{N}$ as a function of applied orienting field.
}
\label{fig:procedure_Ising_XY}
\end{figure}

\begin{figure}
\centering
\includegraphics[width=0.45\textwidth]{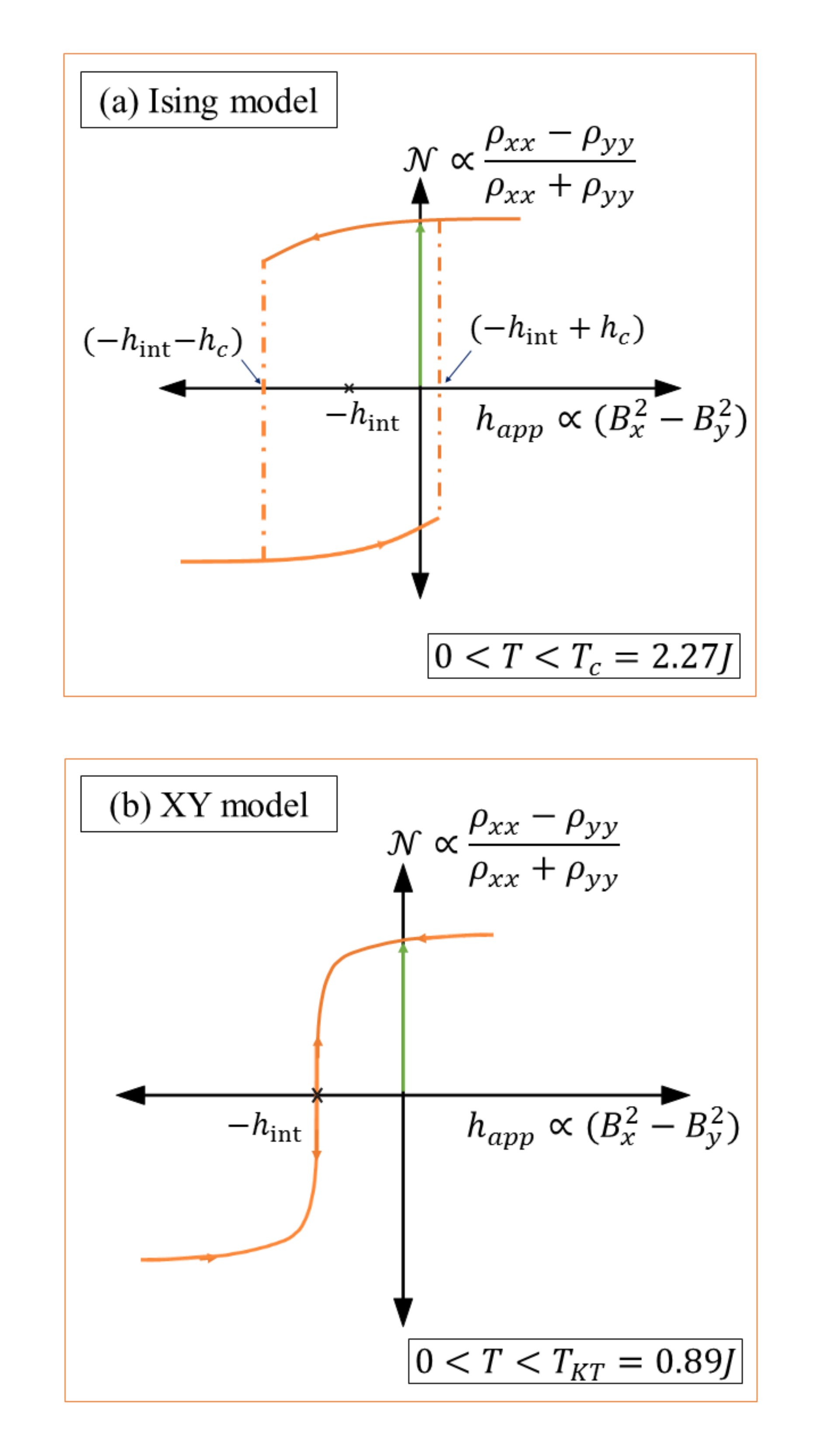}
\caption{
Predicted result of hysteresis test for (a) an Ising nematic and (b) an XY nematic.  
Cooling (green arrow) the system below $T_c$ (Ising) or $T_{KT}$ (XY) 
gives rise to a net nematicity in the presence of any orienting field $h$, including the
case of no applied orienting field, since then $h = h_{int} \ne 0$.
Subsequently sweeping the in-plane orienting field 
gives rise to either hysteresis in the Ising case, or no hysteresis in the XY case.  
}
\label{fig:XY_vs_Ising-hysteresis}
\end{figure}

{\em Experimental test of universality class:}  
We propose that low temperature hysteresis measurements 
can distinguish whether any electron nematic
(including the quantum Hall nematic at high fractional 
filling discussed here) is in the Ising or XY universality class.
Fig.~\ref{fig:procedure_Ising_XY} shows the equilibrium phase diagram for both models,
as a function of temperature and total {\em orienting field} $h$.
Note that because the nematic order parameter switches sign
upon rotating 90$^o$, the orienting field is related to the
applied in-plane magnetic field by $h \propto B_x^2 - B_y^2$.
\cite{Carlson2011}
Other external perturbations also contribute
to an orienting field, such as strain.\cite{Carlson2011,Koduvayur:2011}
In both models, a phase transition only exists 
at zero orienting field, $h = 0$.  
In the Ising case,
the phase transition is into a low-temperature, long-range ordered nematic phase
which spontaneously breaks rotational symmetry.
For the 2D XY model, the phase transition is in the BKT universality class,
and the low temperature phase is critical throughout the temperature range,
with no long range order, and therefore no net nematicity $\mathcal{N}$,
measurable by $\mathcal{N} \propto (\rho_{xx}-\rho_{yy})/(\rho_{xx}+\rho_{yy})$.  
Upon field cooling in any weak $h$, 
{\em both models} will develop a net nematicity below a crossover
temperature which is close to the phase transition temperature,
whether $T_c = 2.27 J$ in the Ising case, or $T_{KT} = .89 J$ in the XY case.

However, hysteresis can clearly distinguish between these universality classes.  
The hysteresis protocol we propose (shown in Fig. \ref{fig:procedure_Ising_XY}) is the following: 
Cool in an orienting field $h>0$ such as in-plane magnetic field (see Ref.~\onlinecite{Carlson2011,KenBCooperThesis} for a list of orienting fields), and go to 
low temperature, well within the nematic region.
Then, reduce $h$ to zero, and sweep it to negative values $h < 0$.  
Using, {\em e.g.}, in-plane magnetic field as an orienting field,
this is equivalent to cooling
with an in-plane field configuration
of $\vec{B}_{\rm in-plane} = (B_x > 0, B_y = 0)$, 
then holding the temperature fixed, 
decreasing $B_x$ to zero, 
then immediately increasing
the field $B_y$ from zero while holding $B_x = 0$
so as to end with an in-plane field configuration
of $\vec{B}_{\rm in-plane} = (B_x = 0, B_y > 0)$.
Indeed, quantum Hall stripes can be reoriented via
application of in-plane field.\cite{Shi:2016}
At low temperature in the Ising case, there is hysteresis in
the net nematicity $\mathcal{N}$ as the in-plane field is swept so as to
take $h$ from positive to negative and back again,
or {\em vice versa}.  
Therefore in the Ising case, the net nematicity should remain in an oriented state,
until the coercive field strength $h_c \ne 0$ is reached.

However, in the XY case, there should be no hysteresis.
This follows from the Mermin-Wagner-Hohenberg theorem, since 
decreasing an applied field $h$ so as to end on the critical phase
at $h = 0$ can leave no long range order, 
$\mathcal{N}(h\rightarrow 0) \rightarrow 0$ where $\mathcal{N}$ is the net nematicity.   
Because $h\rightarrow 0$ with $T < T_{\rm KT}$ is critical, 
$\mathcal{N} \propto h^{(1/\delta)}$ as field is swept, where 
the critical exponent 
$\delta = (4/\eta)-1$ varies from 
$\delta(T_{\rm KT}) = 15$ to $\delta(T\rightarrow 0) \rightarrow \infty$.\cite{Kosterlitz:1974} 
This case is shown in Fig. \ref{fig:XY_vs_Ising-hysteresis}(b). 

It should also be noted that the test is clearest in clean samples,
since addition of random field effects in the presence of a
net orienting field $h$ puts both models in the
universality class of the random field Ising model,\cite{daSilveira:1999}
which has hysteresis at low temperature.  Whereas hysteresis of a clean Ising model
has a net macroscopic jump in the nematicity,
hysteresis of a   random field
Ising model is smooth in two dimensions.\cite{PhysRevB.53.14872}   
At very weak but finite random field strength,
the model predicts avalanches in the resistivity anisotropy around the hysteresis loop
with power law behavior set by critical exponents
characteristic of the 2D random field Ising model critical point.

Note that our simulations as well as those of Ref.~\onlinecite{Nho-2000-PhysRevLett.84.1982} indicate the 
presence of a weak intrinsic orienting field, $h_{\rm int}$ in the sample, on the order 
of $h_{\rm int} \approx 3-5$mK.
This means that to achieve
$h=0$ requires that some extrinsic orienting field, such as an in-plane magnetic field or uniaxial strain,\cite{Carlson2011,KenBCooperThesis}
must be applied to compensate.
Assuming this could be achieved, then zero-field cooling (ZFC) with $h=h_{int}+h_{app} = 0$
has stark differences in the two models.
In the Ising case, ZFC gives rise to long-range order with net nematicity
and macroscopic resistivity anisotropy,
with Ising critical behavior at the onset of nematicity, 
and the direction of that nematicity can randomly switch upon repeated cooling at $h=0$.
In the XY case, ZFC can't produce long-range order or net nematicity,
but the system would instead enter a topological phase with power-law 
nematic order, and accompanying critical phenomena.

In conclusion, we have shown that the entire temperature dependence of the
observed resistivity anisotropy in a high fractional Landau level can be
well described by taking into account the discrete rotational symmetry
of the underlying crystal.  Inclusion of such a symmetry-breaking term
shifts the universality class of the electron nematic from the 
Kosterlitz-Thouless universality class of the two-dimensional XY model
to the two-dimensional Ising universality class.
We furthermore propose an experimental test for hysteresis that can
clearly distinguish whether any 2D electron nematic
is in the Ising or XY (Kosterlitz-Thouless) universality class.

\begin{acknowledgments}
We acknowledge helpful conversations with G.~A.~Csathy, K.~A.~Dahmen,
E.~Fradkin, S.~A.~Kivelson, P.~Muzikar, and H.~Nakanishi.
E.W.C. and S.B. acknowledge support from NSF Grant No. DMR-1508236, and Department of Education Grant No. P116F140459. 
This research was supported in part through computational resources provided by Information Technology at Purdue, West Lafayette, Indiana.
\end{acknowledgments}

\bibliography{QuantumHallNematic-XY-IsingNotes}

\end{document}


\title{Supplementary Information for Distinguishing XY from Ising Electron Nematics}

\author{S. Basak}
\affiliation{Department of Physics, Purdue University, West Lafayette, IN 47907, USA}
\author{E.~W. Carlson}
\affiliation{Department of Physics, Purdue University, West Lafayette, IN 47907, USA}

\date{\today}

\maketitle

\subsection{XY Model with Four-Fold Symmetry Breaking Term $V$}

Inclusion of a four-fold symmetry breaking term in the XY model
yields the following Hamiltonian in the presence of a uniform
orienting field $h$
\begin{eqnarray}
H=&-&J\sum_{\langle i,j \rangle}\cos{(2(\theta_{i}-\theta_{j}))}-h\sum_{i}\cos{(2(\theta_{i}-\phi))} \nonumber \\
&-& V\sum_{i}\cos{(4\theta_{i})}~.
\label{eqn:model_V4}
\end{eqnarray}

\begin{figure}
\centering
\subfigure[\ ]{\label{fig:XY-simulation_a}\includegraphics[width=0.48\textwidth]{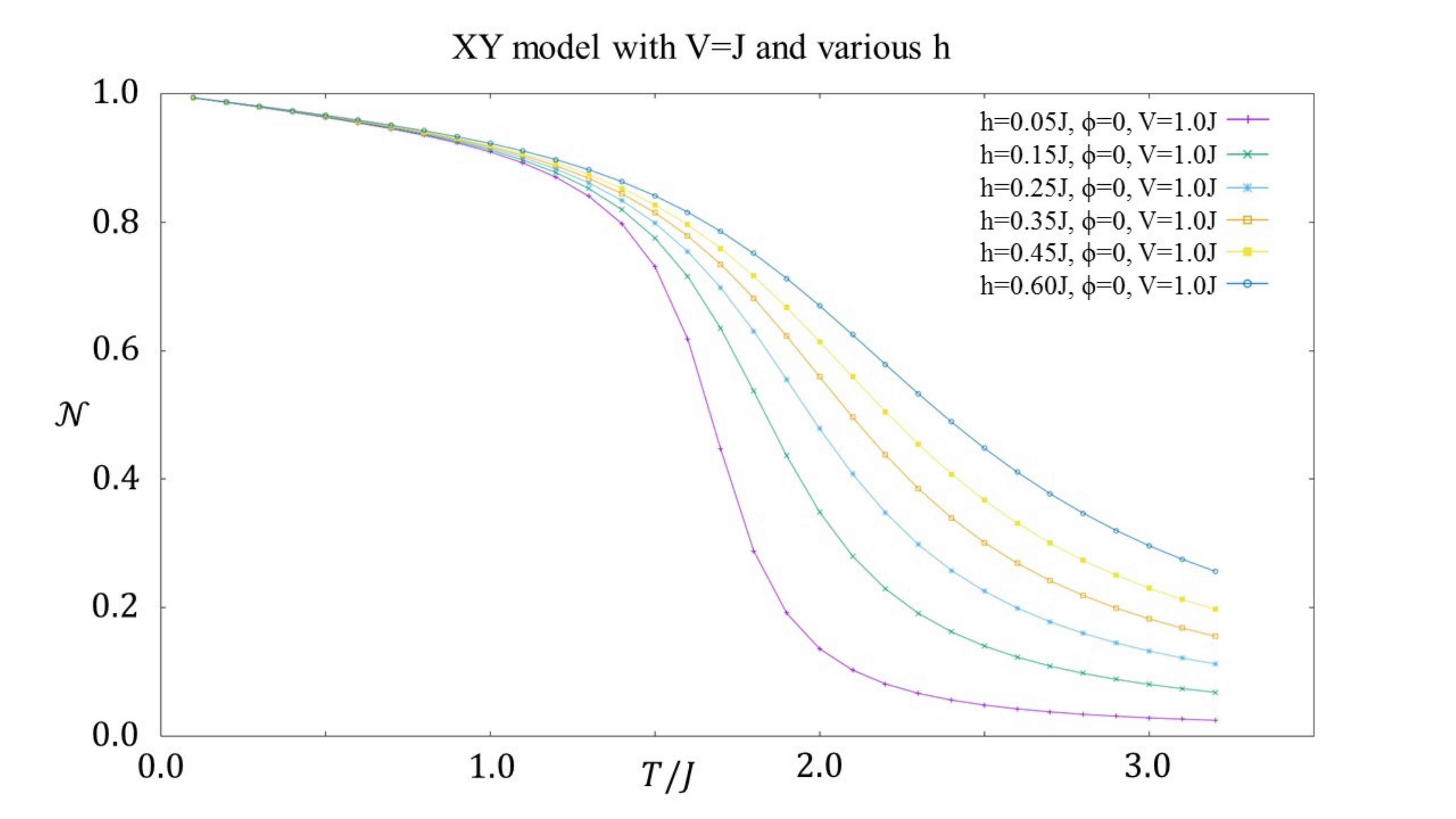}}
\subfigure[\ ]{\label{fig:XY-simulation_b}\includegraphics[width=0.48\textwidth]{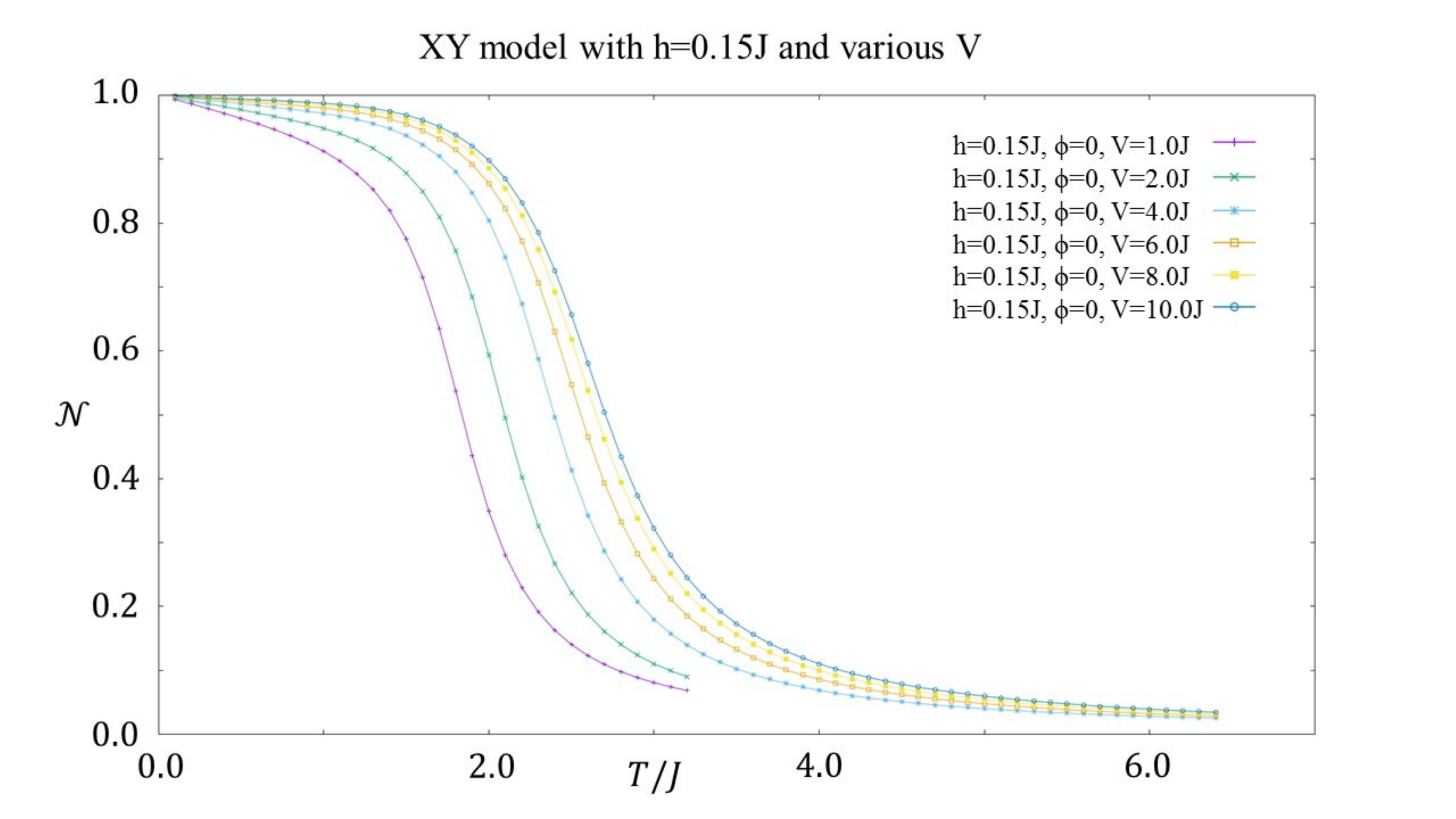}}
\caption{Monte Carlo simulations on a lattice of 100x100 sites,
(a, b) an XY model with a moderate four-fold symmetry breaking field $V$ and uniform orienting field $h$, and 
Note that within an XY description, a moderate 4-fold symmetry breaking term $V \ne 0$ is required to capture the low temperature dependence of the resistivity anisotropy, 
which changes the universality class of the electron nematic from XY to Ising.    
}
\label{fig:XY-simulation}
\end{figure}

Fig. \ref{fig:XY-simulation} shows how different values of $V$ and $h$ change the behavior of $\mathcal{N}$ as a function of temperature.
Increasing $h$ from zero has little effect on the low temperature behavior, but it rounds the transition, leading to an 
increasingly high temperature for the onset of nematicity, as shown in Fig.~\ref{fig:XY-simulation_a}.
Increasing $V$ from zero causes a change in the low temperature slope of the nematicity,
flattening the curve at low temperature, as shown in Fig.~\ref{fig:XY-simulation_b}. Inclusion of the four-fold symmetry breaking term $V$ also
shifts the transition temperature from the XY value $T_{\rm KT} = .89 J$ towards the Ising value
$T_c = 2.27 J$.

\subsubsection{Effect of Changing $\phi$}

\begin{figure}
\centering
\subfigure[\ $\phi=0$]{\label{fig:site_potential-1}\includegraphics[width=0.48\textwidth]{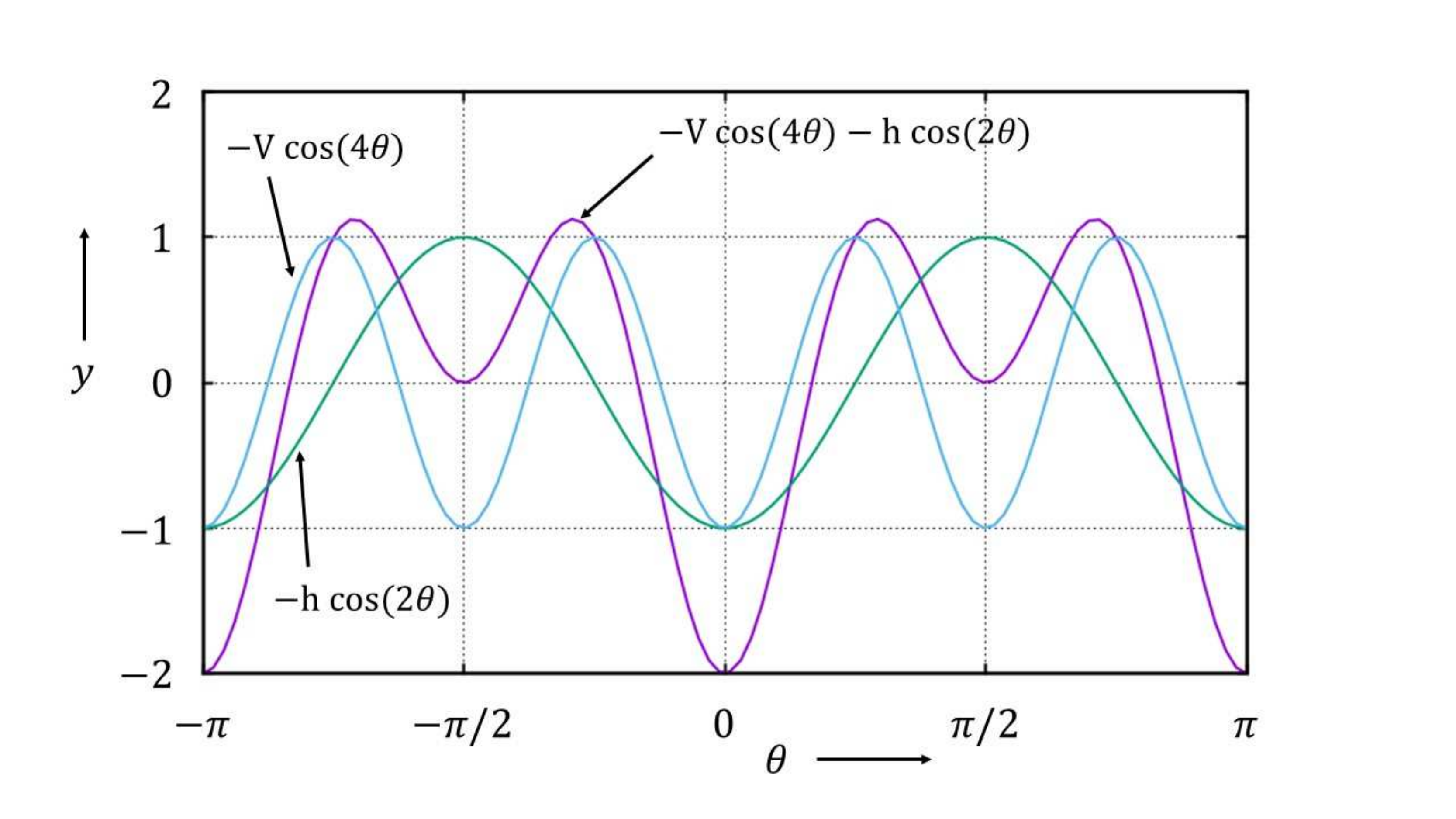}}
\subfigure[\ $\phi=\pi/12$]{\label{fig:site_potential-2}\includegraphics[width=0.48\textwidth]{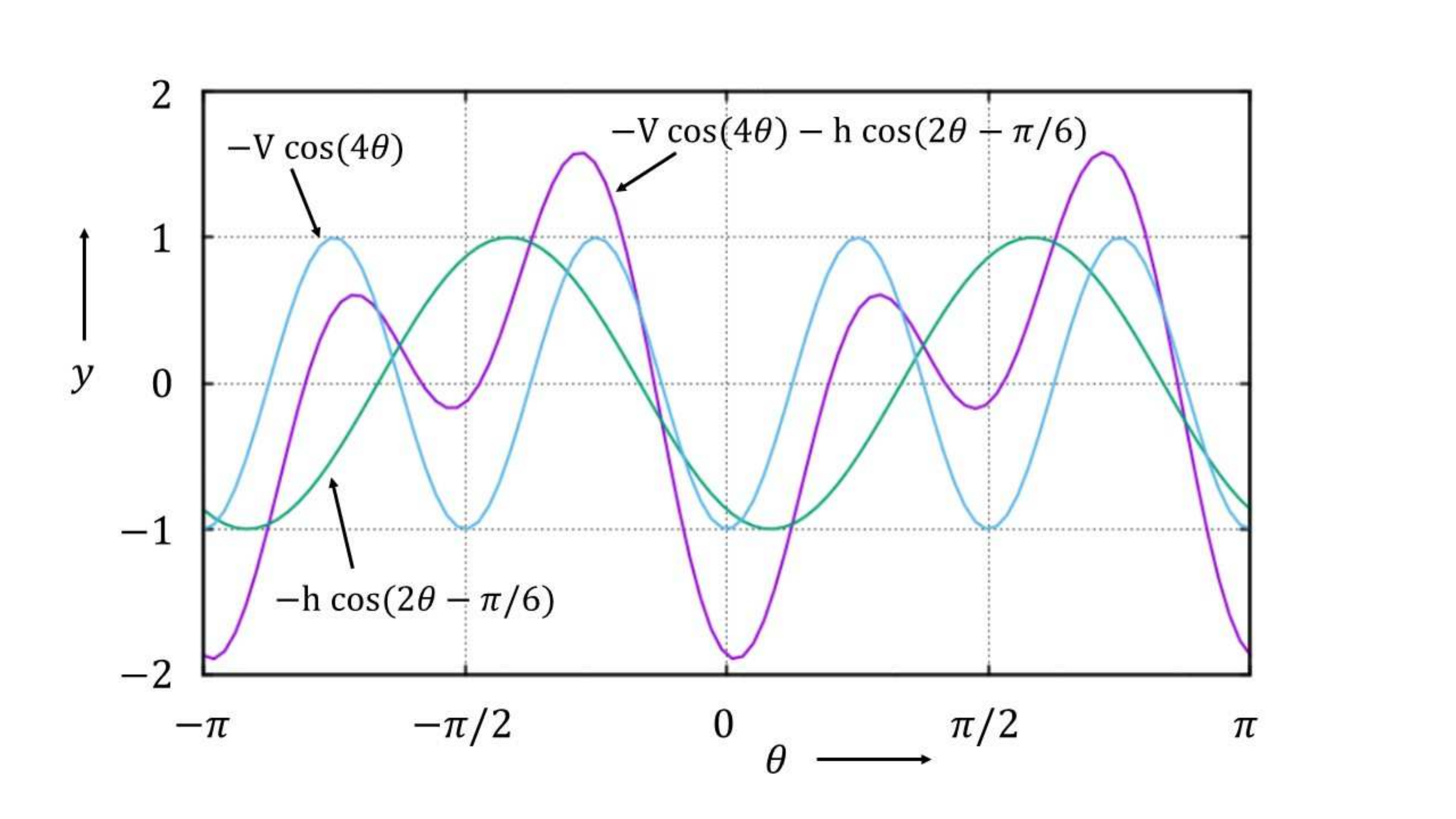}}
\subfigure[\ $\phi=\pi/6$]{\label{fig:site_potential-3}\includegraphics[width=0.48\textwidth]{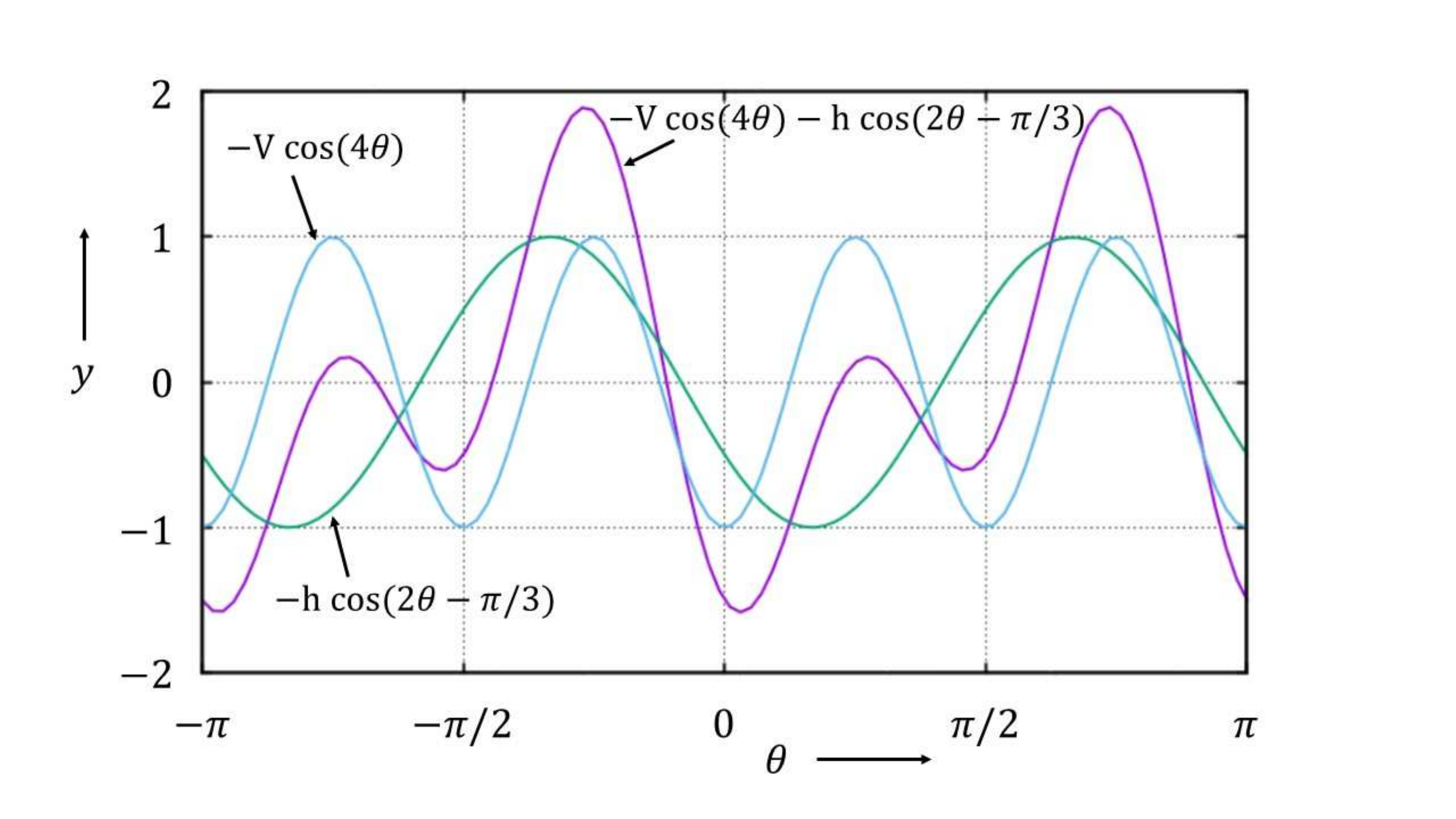}}
\subfigure[\ $\phi=\pi/4$.]{\label{fig:site_potential-4}\includegraphics[width=0.48\textwidth]{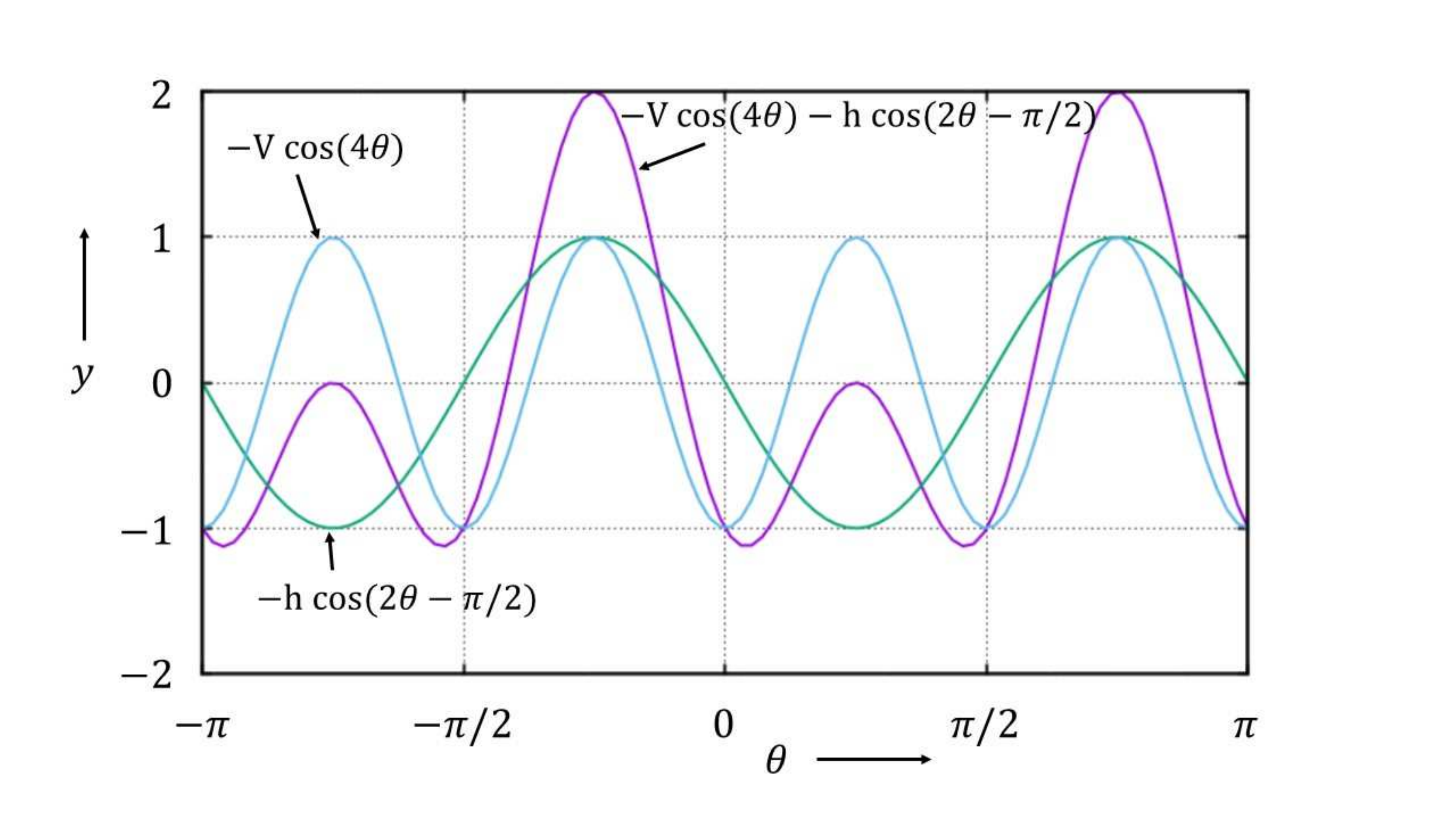}}
\caption{Plot of a single site potential energy as a function of $\theta$. The $h$-term with different $\phi$ is plotted in green, the $V$-term is plotted in blue and the total site energy ($V$-term+$h$-term) is plotted in purple. (a) The difference between the 2 distinct potential minima is largest, (d) All the potential minima are at same energy.}
\label{fig:site_potential}
\end{figure}

Whereas the four-fold symmetry breaking term $V$ describes
the fact that the electron nematic tends to lock into a major crystalline axis direction,
applying a uniform orienting field $h$ can disrupt this, if the angle
$\phi$ between the direction of the orienting field and the crystalline axes is large enough.  
When the orienting field $h$ is aligned with the crystal, $\phi = 0$, and 
the ground state configuration  is $\left< \theta_{i} \right>=0$  $\forall$  
i  (lattice points). 

Fig.~\ref{fig:site_potential} shows the on-site potential energy $-V \cos{(4\theta)}-h \cos{(2(\theta-\phi)})$ with $V=h=1$.
This figure illustrates the effect of $V$ and $h$
on the single site potential energy of 
Eqn.~\ref{eqn:model_V4} as the angle $\phi$ is varied.
For $\phi = 0$, the uniform orienting field $h$ is aligned with a
major crystalline axis, which explicitly breaks the rotational symmetry. 
This further reinforces the tendency of the nematic to lock to a crystalline axis
and at low temperature, the average value of $\theta_i$ localizes
in a global minimum of the purple curve in Fig.~\ref{fig:site_potential}(a),
$\left< \theta_{i} \right>=0$.  

Because the term $V\cos{(4\theta_i)}$ forms a steeper well near its minimum than does 
the term $h\cos{(2(\theta-\phi))}$,
the addition of the $V$ term suppresses the low temperature
orientational fluctuations of the  nematic.
This is the origin of the flatness of the nematicity {\em vs} temperature for 
$V$ large compared to $J$.

The crystal field term $V$ breaks the rotational symmetry of the system from $C_{\infty}$ to $C_4$ and the $h$-field breaks the $C_4$ symmetry to $C_2$ symmetry. The orientation of $h$-field at an angle $\phi=45^o$ gives rise to rectangular
symmetry instead of square ($C_4$) symmetry. 
As the angle ($\phi$) the $h$-field makes with the X-axis (See Fig. 1 of main text) is increased from $0^o$ to $45^o$, the difference in the depth of the wells decreases to zero (See Fig. \ref{fig:site_potential}). Hence the transition region of $ \mathcal{N}(T) $ becomes steeper as $\phi$ goes from $0^o$ to $45^o$ (See Fig. 3 of main text). 

In the configuration $\phi=45^o$, the orienting field $h$ no longer 
favors one minimum of the $V$ term over the other minimum.
Therefore, at small $h<4V$ with $\phi =45^o$, the low temperature
phase spontaneously breaks the symmetry between the two minima of 
the $V$ term, and the transition is sharp, as shown in Fig.~1 of the main text.  
However, for $h\ge4V$, even with $\phi=45^o$,
the low temperature phase no longer spontaneously breaks symmetry,
and the transition is rounded.

\subsection{Ising model}

\begin{equation}
H=-J\sum_{\langle i,j \rangle}\sigma_{i} \sigma_{j}-h\sum_{j}  \sigma_{i}~.
\label{eqn:model_ising}
\end{equation}

\begin{figure}
\centering
\includegraphics[width=0.48\textwidth]{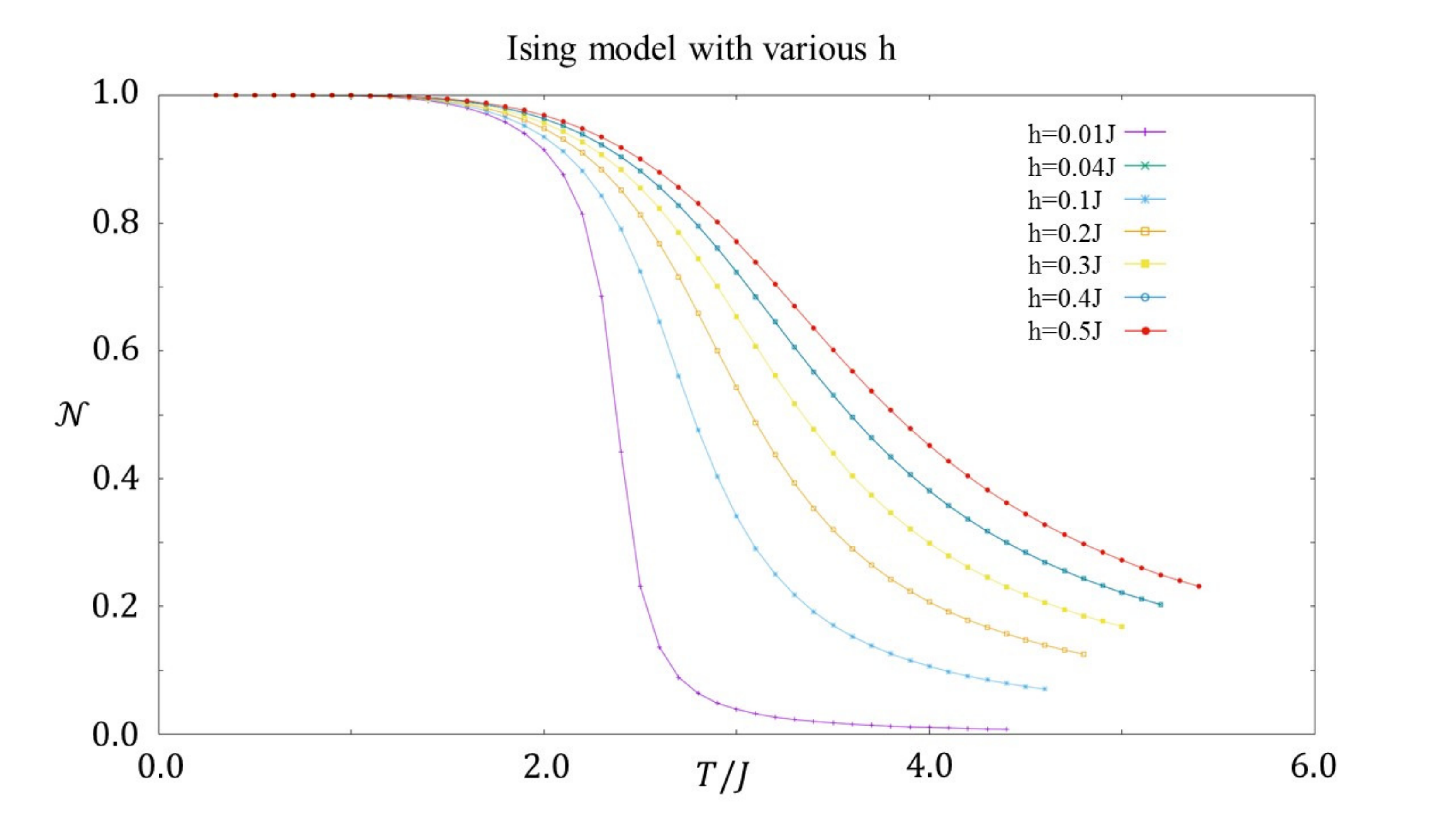}
\caption{Monte Carlo simulations on a lattice of 100x100 sites,
of an Ising model with uniform orienting field $h$.  
Note that within an Ising description, for larger values of $h$ the nematicity gets rounded more towards the high temperature isotropic phase.    
}
\label{fig:Ising-simulation}
\end{figure}

Fig. \ref{fig:Ising-simulation} shows how different values of $h$ change the behavior of $\mathcal{N}$ as a function of temperature.
Increasing $h$ from zero rounds the transition, leading to an 
increasingly high temperature for the onset of nematicity. 
It has similar effects as increasing $h$ in the XY model with a constant $V$ (Compare with Fig.~\ref{fig:XY-simulation_a}).

\subsection{\em Intrinsic orienting field: Origin of $h_{\rm int}$}
There are a number of possibilities for the origin of the intrinsic orienting field $h_{\rm int}$,
as discussed extensively in Ref.~\onlinecite{KenBCooperThesis},
such as a miscut, MBE growth imperfections, or slight misalignment of the applied perpendicular magnetic field. 
In a AlGaAs heterojunction, the chemical bonds link up at the junction in either the [$110$] or the [$1\bar{1}0$] orientation, {\em i.e.} bonds are stretched slightly more along one of the crystalline axes than the other due to lattice constant mismatch. 
(Note that the lattice constant $a=(5.6533+0.0078{\rm x})$\AA \ at 300K, where $x$ is the Al doping in ${\rm Al_{x}Ga_{1-x}As}$.\cite{footnote-AlGaAs}). 
This gives rise to uniaxial physical strain in either the [$1\bar{1}0 $] or [$110$] directions. 
This also means that the random lattice points occupied by Al at the interface 
are a possible source of random field effects.\cite{footnote-disorder_effects}

\bibliography{supplementary_info}